# Large Language Model-Assisted Discovery of Optimal Dopants for Enhanced Thermoelectric Performance in CoSb$_3$-Based Skutterudites

Yagnik Bandyopadhyay,[1] Dylan Noel Serrao,[1] and Houlong L. Zhuang[1,*]

[1]School for Engineering of Matter Transport and Energy, Arizona State University, Tempe, AZ 85287, USA

*zhuanghl@asu.edu

**Abstract**

We present a data-driven approach for accelerating the discovery of high-performance $CoSb_3$-based skutterudites by curating a comprehensive dataset of compositions with various filler elements from over 300 research articles. Leveraging large language models (LLMs), we extract and embed compositional representations, which are then used to train a regression head for predicting thermoelectric figure of merit. Compared to traditional deep neural networks relying on elemental descriptors such as atomic radii, our LLM-based model achieves significantly lower mean-squared error losses. We further employ the trained model to propose novel filler compositions with promising thermoelectric properties. Finally, we support these predicted candidates through density functional theory and molecular dynamics calculations to assess their electrical and thermal conductivity. This data-driven approach demonstrates the potential of combining natural language processing, machine learning, and quantum simulations for thermoelectric materials design.

# 1. Introduction:

Growing global energy demand and the depletion of fossil fuel resources highlight the urgent need for cleaner, more sustainable energy solutions [1,2]. Thermoelectric materials have emerged as promising candidates for clean energy solutions by converting waste heat into electrical power. These materials are unique in that they can directly transform thermal energy into electrical energy through the Seebeck effect, and electrical energy into thermal energy through the Peltier effect [3–6]. The performance of thermoelectric materials is quantified by the dimensionless figure of merit, ZT,

$$ZT = \frac{S^2 \sigma T}{\kappa}$$

where S is the Seebeck coefficient, σ is the electrical conductivity, T is the absolute temperature, and κ is the thermal conductivity [7–9]. A high ZT value generally indicates higher electrical conductivity and lower thermal conductivity.

The development of thermoelectric materials with Phonon-Glass Electric Crystals (PGEC) behavior offers a transformative opportunity to enhance energy conversion efficiency by decoupling thermal and electrical transport properties [10]. Materials such as skutterudites (e.g., $CoSb_3$) [11], clathrates (e.g., $Sr_8Ga_{16}Ge_{30}$) [12,13], and half-Heusler alloys like (Hf, Zr) NiSn [4] exemplify this behavior through structural mechanisms that scatter heat-carrying phonons while preserving efficient electron transport. For example, skutterudites exhibit a distinctive atomic structure containing voids enclosed by octahedral motifs located at the corners of a cubic unit cell. These void sites can also be interpreted as lattice points within a body-centered cubic cell. The voids provide opportunities for external atoms, often referred to as "rattlers," to occupy them, thereby inducing structural distortions and charge-transfer effects that significantly influence the electronic band structure and phonon-scattering mechanisms [6,11]. Filler atoms such as Ba, Yb, and In have been shown to enhance thermoelectric performance by increasing carrier concentration and reducing lattice thermal conductivity, although their specific effects on electrical conductivity remains underexplored [14]. In addition, these filler atoms act like Einstein oscillators, moving relatively freely within the voids and causing resonant scattering that reduces the phonon mean free path [15]. This effect may enable the thermal conductivity of skutterudites to approach the minimum thermal conductivity limit observed in glasses. Similarly, other PGEC materials like layered cobaltates (e.g., $(Ca_2CoO_3)_{0.62}(CoO_2)$) [16] and doped compounds such as $AgSbTe_2$ with Yb form nanoscale superstructures or exhibit vibrational disorder that enhance phonon scattering while maintaining high electron mobility [17]. These materials underscore the potential of engineering atomic-scale and nanoscale features to achieve the ideal PGEC paradigm, thereby paving the way for advances in thermoelectric technologies for sustainable energy applications.

This work focuses on skutterudites, one of the most promising families of phonon-glass electron-crystal (PGEC) thermoelectric materials, using $CoSb_3$ as a representative example to illustrate our research objectives without loss of generality. $CoSb_3$ exhibits favorable electrical transport properties, but in its pristine form it is not well suited for high-performance thermoelectric applications because of its thermal transport characteristics and the absence of filler atoms in its voids. When these voids are filled with external atoms, $CoSb_3$-based skutterudites can efficiently convert heat into electricity and typically exhibit thermoelectric ZT values greater than 1.0 [18,19]. A wide range of elements from the alkali-earth and lanthanide groups have been used as filler atoms in $CoSb_3$-based skutterudites [20–23]. The type and concentration of these filler atoms play crucial roles in affecting the electrical and thermal conductivity of the material. While skutterudites with single, binary, or ternary filler atoms have been studied efficiently, a deeper fundamental understanding is needed to guide the rational design of these thermoelectric materials and to overcome the challenge of simultaneously achieving high electrical conductivity, low thermal conductivity, and a high ZT value [13,24–26].

Traditionally, the discovery of thermoelectric materials relied on experimental trial and error method combined with computational techniques like density functional theory (DFT) and molecular dynamics (MD) [27–29]. These computational techniques are generally very expensive in terms of time and computational resources [30]. This limits to screen through the vast chemical and compositional space of thermoelectric materials [31–33]. This challenge has prompted a growing interest in data-driven and machine learning (ML) techniques, which aim to accelerate materials discovery by leveraging existing data to predict thermoelectric properties, thereby reducing dependence on resource-heavy first-principles simulations [27,34–36].

Artificial neural networks (ANNs) [37,38] have gained popularity as forward prediction models for estimating thermoelectric properties from material descriptors [35,36,39]. These models can capture complex nonlinear relationships among composition, structure, and performance metrics such as the Seebeck coefficient and ZT. However, ANNs typically require input features derived from physical properties, such as elemental attributes or structural parameters, which are not always available or consistently reported in existing datasets.

Large language models (LLMs) [40], such as GPT [41] have revolutionized natural language processing and are increasingly being applied in scientific domains to extract and encode complex information from unstructured data [9,42,43]. Inspired by this capability, our approach leverages LLMs to embed compositional information of thermoelectric materials without requiring explicit structural input and use these embeddings as inputs to a neural-network-based forward model. Although prior studies have incorporated LLM-based features for property prediction, they typically rely on additional crystal-structure information or domain-specific featurization [44]. In contrast, our framework is designed to predict the ZT using only the raw compositional formula. This approach enables us to screen and propose new $CoSb_3$-based skutterudites with promising thermoelectric performance, even in the absence of detailed structural data.

In this work, we curate a focused dataset of $CoSb_3$-based skutterudites by manually collecting composition and ZT value from nearly 300 published research articles. We then organize this information into a format compatible with language-model training, similar in style to instruction-based datasets such as Alpaca [45–47], which allows us to treat each material composition as a natural language input. Using this setup, we embed the compositions using a pretrained BERT [48] model. These embeddings are then passed through a regression head to predict ZT value using only the composition and temperature, enabling a fully data-driven forward model to predict the ZT values of random $CoSb_3$-based skutterudite compositions. We next employ a random-sampling-based skutterudite generator to propose candidate compositions and screen them by selecting those with low (< 0.5) and high (> 1.0) predicted ZT values. Finally, we validate the model predictions through density functional theory (DFT) and molecular dynamics (MD) simulations to estimate the electrical and thermal conductivities of the selected skutterudite candidates.

## 2. Methods

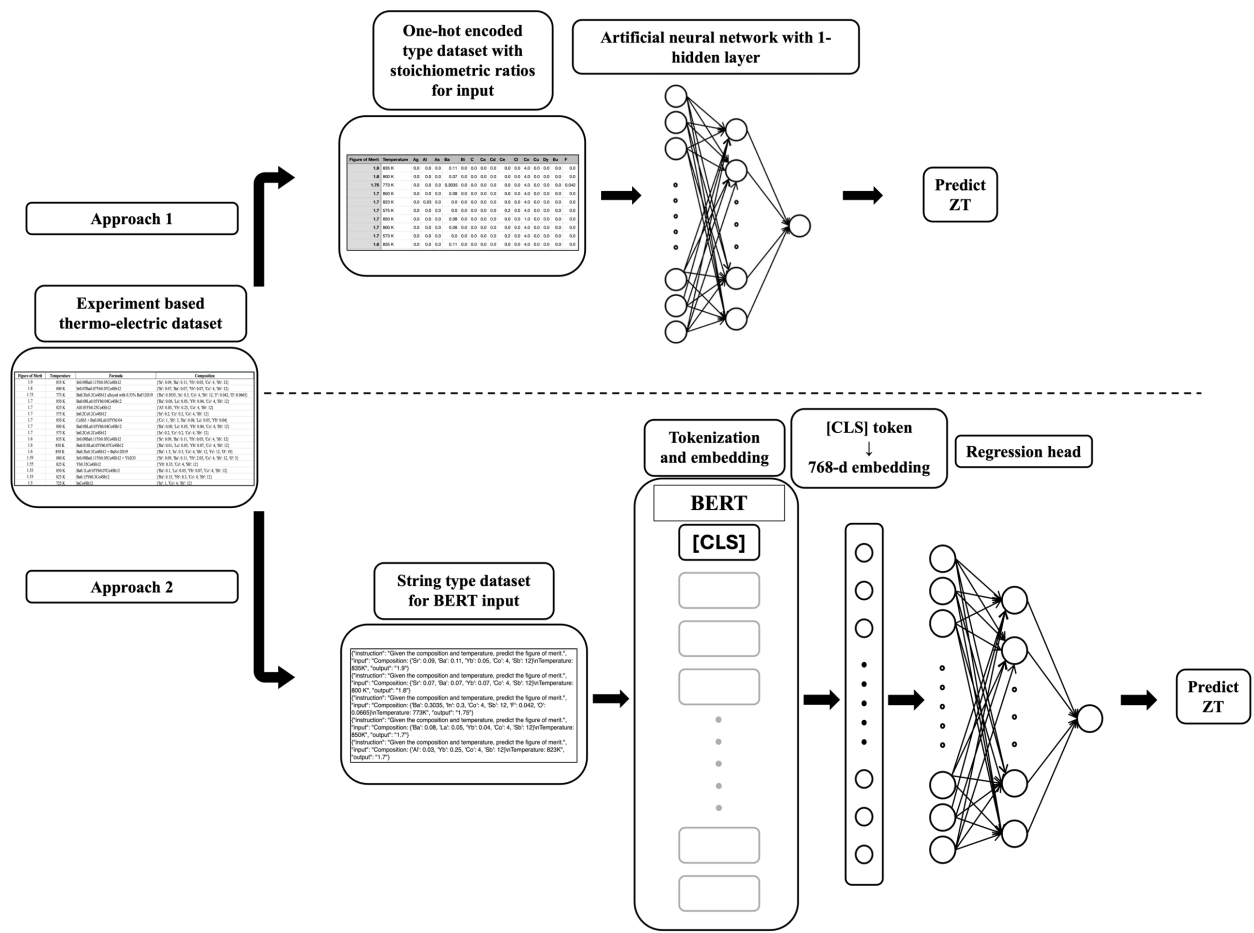

**Figure 1:** Pipeline of the two approaches to predict ZT values. Approach 1 is a forward model using an artificial neural network, while approach 2 is a BERT-based model with a regression head for predicting ZT values.

**Figure 1** illustrates the dataset construction and modeling workflow adopted in this study. We start with curating a focused dataset consisting of 412 experimental results for $CoSb_3/Co_4Sb_{12}$-based skutterudites, extracted from over 300 research articles. For each entry, we record the chemical composition, measurement temperature, and corresponding ZT value. We explore two modeling approaches: (1) a baseline forward model using an ANN and (2) an LLM-based approach using BERT to embed the input compositions. The BERT embeddings are then fed into a simple regression head with a single hidden layer to predict the ZT values.

Both models are trained to take composition and temperature as input and output the predicted ZT. We compare their performance to evaluate how much the BERT-based embedding improves over the conventional ANN model. Finally, we apply a random sampling strategy to generate new compositions of CoSb$_3$-based materials with single, binary, and ternary filler elements. The ZT values of these compositions are predicted using the LLM-based model, and the most promising candidates are further validated using DFT and MD simulations.

## 2.1 LLM and ANN Training

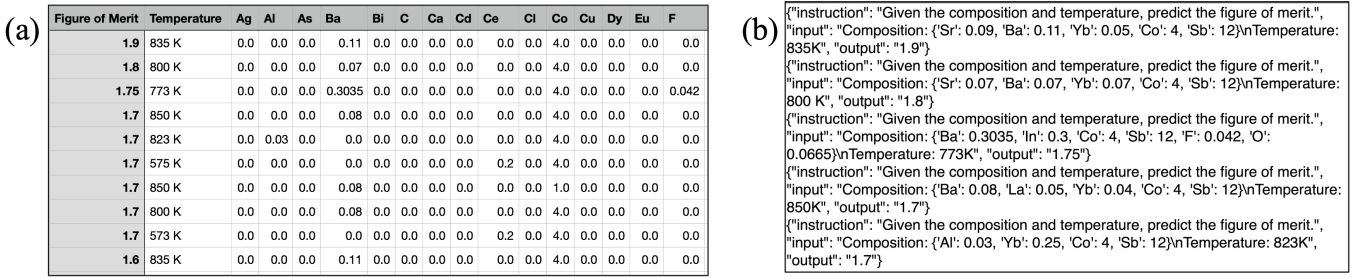

**Figure 2:** (a) One-hot encoded style representation of the composition used in the ANN model, (b) String-type dataset for input to the BERT-based model.

To predict the thermoelectric ZT from material composition and temperature, we use two different modeling approaches. The first is a conventional ANN, and the second leverages a LLM to generate embeddings from composition and temperature strings.

In the ANN-based approach, each composition is converted into a one-hot encoded vector that captures both the presence of different elements and their stoichiometric ratios as shown in figure 2(a). This vector, along with the temperature value as a separate numerical input, is fed into a fully connected neural network with two hidden layers and ReLU activation functions. The model is trained using mean squared error (MSE) loss and optimized with the Adam optimizer. In the LLM-based approach, we treat the composition as a natural language input and format our dataset in the style of Alpaca-style instruction[45–47], where each input consists of a textual composition and temperature string as shown in Figure 2(b). We use a pre-trained BERT model ("bert-base-uncased") to tokenize and embed these inputs. Specifically, we extract the "[CLS]" token embedding from the final hidden layer of the model, resulting in a 768-dimensional vector that represents the entire input. To ensure that the model is invariant to the order of elements in the composition string, we remove the positional encoding [49] from BERT by setting the positional embedding weights to zero. This ensures that permuting the order of elements in the input does not affect the resulting embedding.

The BERT-generated embedding is then passed through a lightweight regression model with one hidden layer to predict the ZT value. Both the ANN and the LLM-based regressor are trained to take composition and temperature as input and produce the ZT as output. To ensure consistency and robustness, we train each model using 30 seeds generated randomly, then use the same seeds for both training and inference. This allows us to capture general trends and assess the variability in model predictions. All models are implemented in PyTorch[50], and the BERT-based model is accessed using the Hugging Face Transformers library [51]. Training and evaluation are conducted on GPU-enabled machines to reduce computational time.

**2.2 Random Sampling**

To identify novel filler combinations with promising thermoelectric performance, we use a random sampling strategy guided by our trained regression model. The goal is to explore the composition space of $CoSb_3$-based skutterudites by stochastically generating candidate compositions and predicting their corresponding ZT value.

We define a sampling space consisting of possible filler elements commonly used in skutterudite systems, including both alkaline earth and rare earth elements (e.g., Ba, Yb, La, Ce, Sr, etc.). Each composition is constructed by randomly selecting one or more filler elements and assigning stoichiometric ratios such that the total occupancy does not exceed the available void sites in the $CoSb_3$ supercell. The random generation process ensures physical validity by enforcing constraints on the total number of atoms and avoiding unrealistic occupancy distributions.

Once a random composition is generated, it is formatted as a text input in the same style used during LLM training and passed through the embedding and regression model. The trained model then predicts the FOM at a temperature value selected randomly from a predefined range.

To increase the diversity of the sampling pool, we generate and evaluate thousands of such compositions. Only those that exceed a specified FOM threshold (e.g., FOM > 1.0) are retained for further analysis. This filtering helps focus on potentially high-performing candidates while minimizing the computational cost of subsequent ab initio validation.

Additionally, temperature-dependent sampling is performed by pairing each composition with a randomly selected temperature within a defined range (e.g., 300 K to 800 K). This approach provides a reasonable estimate of how the ZT values for the same filler atoms vary under different thermal conditions.

## 2.3 DFT Calculations

To validate our predictions from LLMs, we perform DFT calculations to predict the electrical conductivity of supercell models. The Kubo-Greenwood[52,53] formula is a foundational theoretical framework for determining electrical conductivity in materials, derived from linear response theory. It establishes a direct relationship between the electronic structure and transport properties by incorporating velocity matrix elements and energy differences between occupied and unoccupied states. This formalism enables precise calculations of conductivity across diverse material systems, including crystalline solids[54], disordered systems[55], high-entropy alloys[56], and amorphous semiconductors[57]. Its versatility and proven success in handling complex electronic interactions make it particularly suitable for analyzing filled $CoSb_3$, where intricate interactions between host lattice atoms and filler atoms significantly influence transport properties. A key advantage of the Kubo-Greenwood formalism is its ability to account for finite-temperature effects and scattering phenomena, critical factors for realistic modeling of thermoelectric materials like $CoSb_3$. These materials exhibit temperature-dependent behavior that strongly impacts their performance. By summing over electronic transitions between energy states weighted by their respective probabilities and velocity matrix elements, the formalism provides a detailed understanding of how changes in the electronic structure, such as those induced by filler atoms, affect conductivity.

The electrical conductivity of all $CoSb_3$-based systems is obtained using the Kubo-Greenwood formalism as implemented in the *kg4vasp* post-processing package [52,58]. Each structure is first fully relaxed in VASP to obtain its optimized geometry. DFT calculations are performed using the Vienna Ab initio Simulation Package (VASP)[59–61] within the projector augmented-wave (PAW)[62] formalism, with PAW-PBE potentials[63,64] used for all elements. The exchange-correlation functional is described using the generalized gradient approximation in the Perdew-Burke-Ernzerhof (PBE)[65] form. A plane-wave energy cutoff of 600 eV is employed for all calculations. Brillouin-zone integrations are carried out using a 2 × 2 × 2 Monkhorst-Pack k-point mesh for all supercells. Structural relaxations are performed using a conjugate gradient algorithm, with full relaxation of both atomic positions and lattice parameters. The electronic self-consistency convergence criterion is set to $10^{-6}$ eV, and the ionic relaxation is continued until the forces on all atoms are less than 0.01 eV/Å. A Gaussian smearing scheme is used, with smearing widths corresponding to the target temperatures of each system, specifically, $\sigma$ = 0.0587 eV (≈ 682 K) for $CoSb_3$ and $CoSb_3Ce_{0.078125}In_{0.03125}Ba_{0.03125}$, and $\sigma$ = 0.0456 eV (≈529 K) for $CoSb_3Ag_{0.03125}$. A maximum of 300 ionic steps (NSW = 300) is allowed to ensure full structural relaxation.

A static SCF calculation is then performed on the relaxed structure to generate converged eigenvalues and wavefunctions required for transport calculations. The SCF calculations are performed using a plane-wave cutoff energy of 600 eV with a total of 256 bands and employed Gaussian smearing with a smearing width corresponding to the target temperature (e.g., σ = 0.0587 eV for 682 K). Following the SCF calculation, a *nabla* calculation to compute the velocity (momentum-gradient) matrix elements between electronic states. This step is carried out using the LOPTICS flag with a dense energy grid (NEDOS = 2000) and included a total of 256 bands to ensure consistency with the SCF calculation. These quantities serve as the required inputs for *kg4vasp*, which evaluates the frequency-dependent conductivity tensor using the standard Kubo-Greenwood expression adapted for PAW datasets as given in [52]:

$$\sigma(\omega) = i \frac{2e^2 \hbar^3}{m_e^2 V} \sum_{m,m'} \frac{[f(\varepsilon_{m'}) - f(\varepsilon_m)]}{(\varepsilon_m - \varepsilon_{m'})} \frac{\langle m | \nabla | m' \rangle \langle m' | \nabla | m \rangle}{(\varepsilon_m - \varepsilon_{m'} - \hbar\omega + i\delta/2)},$$

where $f(\varepsilon)$ is the Fermi-Dirac occupation, $\langle m | \nabla | m' \rangle$ are the velocity (gradient) matrix elements between Kohn-Sham states, $V$ is the system volume, and $\delta$ is the spectral broadening parameter. The dc electrical conductivity $\sigma_{dc}$ reported in this work corresponds to the $\omega \to 0$ limit of $\sigma(\omega)$ after numerical broadening and evaluation at the target temperature by *kg4vasp*. This workflow provides an *ab initio* estimate of intrinsic electronic transport in pristine and doped $CoSb_3$ supercells.

To determine the lattice thermal conductivity of the three systems, we follow the workflow established by Brorsson et al. [66], which requires training of a force constant potential (FCP) file from rattled structures which is generated from AIMD simulations, the FCP file is then fed into GPUMD simulation which provides the lattice thermal conductivity as the output [67,68]. For each composition, we extract approximately 100 - 250 AIMD snapshots and use them as training data to construct a fourth-order force constant potential file via the Hiphive package [68]. The resulting FCP file is then supplied to GPUMD to perform equilibrium molecular dynamics (EMD) simulations, from which the lattice thermal conductivity is computed using Green-Kubo formalism.

In the GPUMD simulations, a time step of 1 fs is used. The systems are first equilibrated in the canonical (NVT) ensemble using a Nosé-Hoover chain thermostat for 500,000 steps (500 ps). Following equilibration, production runs are carried out in the microcanonical (NVE) ensemble for 10,000,000 steps (10 ns). The heat current autocorrelation function (HAC)[69] is computed during the NVE phase using a sampling interval of 10 steps (10 fs) and a correlation window of 100,000 steps (100 ps).

The thermal conductivity is then obtained by integrating the HAC function according to the Green-Kubo[69–72] relation. The HAC function describes how long the heat current remains correlated with its initial value during an equilibrium MD trajectory. Its decay characterizes the lifetime of heat-carrying phonons: slower decay corresponds to higher thermal conductivity. Using the Green-Kubo relation, $\kappa$ is obtained by integrating the HAC over time until convergence. This approach allows us to capture thermal transport behavior with accuracy close to first-principles methods, making it well suited for comparing thermal conductivity across the different systems.

## 3. Results and Discussion

### 3.1 LLM and ML Training

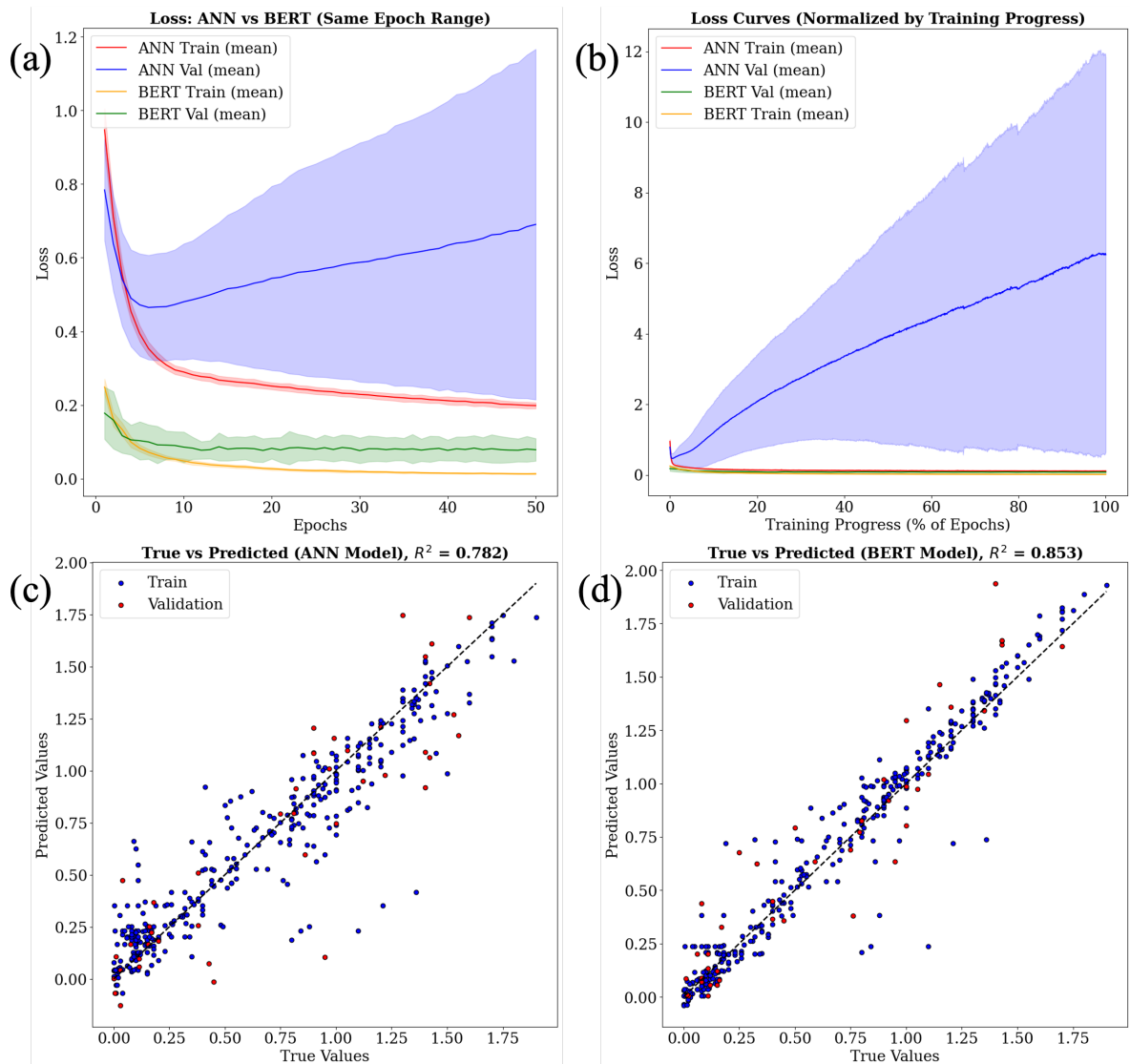

**Figure 3:** (a) MSE loss plots of ANN training and validation as well as BERT based model training and validation for the first 50 epochs, (b) MSE loss plots of both the models normalized by training progress, (c) True v/s Predicted plot of the best ANN model, (d) True v/s Predicted plot of the best BERT based model.

The training behavior and predictive performance of the BERT-based regression model are compared against a baseline artificial neural network (ANN) model in Figure 3. The LLM-based model (BERT) is trained for 50 epochs, as it achieves stable and optimized mean squared error (MSE) losses early in training. In contrast, the ANN model is trained for 1000 epochs to minimize loss further. However, on average, the ANN model still exhibits higher variability and instability across different runs. To enable a fair comparison, we evaluate both models at the 50th epoch. As shown in Figure 3(a), the BERT-based model consistently achieves lower and more stable training and validation losses compared to the ANN model over the same epoch range. When the loss curves are normalized by training progress as shown in Figure 3(b), the ANN validation loss displays a pronounced upward trend with large variance, indicating poor generalization and sensitivity to different seed values for initialization. In contrast, the BERT model maintains a nearly flat and low validation loss, highlighting its robustness with respect to training duration.

The predictive accuracy of the two models is further illustrated through true vs. predicted ZT plots in Figure 3(c–d). The ANN model, as shown in Figure 3(c), exhibits noticeable scatter around the ideal parity line, particularly for higher ZT values, whereas the BERT-based model, shown in Figure 3(d), shows a tighter clustering of predictions along the diagonal, indicating improved regression fidelity. This visual trend is consistent with the higher coefficient of determination obtained by the BERT model.

A quantitative comparison across 30 random seeds is summarized in Table 1. While the best ANN seed achieves a reasonable validation performance, its overall behavior is highly unstable, as reflected by large standard deviations in MSE and coefficient of determination ($R^2$). Notably, the ANN model produces negative mean $R^2$ values across seeds, indicating frequent failure to generalize. In contrast, the BERT-based model demonstrates both lower mean errors and significantly reduced variance, with a best-case validation MSE of 0.0373 and a best $R^2$ of 0.8527. This consistent performance supports the hypothesis that the ANN model suffers from overfitting, likely due to the limited size of our dataset. The BERT-based model, leveraging pretrained language representations of chemical compositions, exhibits superior generalization and stability, making it more suitable for downstream screening tasks.

To further assess the practical utility of the BERT-based model, it is subsequently used to predict ZT values for randomly generated compositions. A selection of these high predicted-ZT candidates is then subjected to DFT and MD simulations to evaluate their thermoelectric stability and validate the model-driven screening results.

**Table 1.** Performance comparison of validation prediction of ZT values for both BERT based model and ANN model.

| Model | Best MAE | Best MSE | Best $R^2$ | Mean MAE | Std MAE | Mean MSE | Std MSE | Mean $R^2$ | Std $R^2$ |
|---|---|---|---|---|---|---|---|---|---|
| BERT | 0.1388 | 0.0373 | 0.8527 | 0.1996 | ± 0.0295 | 0.086 | ± 0.0264 | 0.6433 | ± 0.1247 |
| ANN | 0.1846 | 0.0644 | 0.7824 | 0.4793 | ± 0.201 | 1.8983 | ± 1.7835 | -6.8143 | ± 7.1057 |

## 3.2 Random Sampling

**Table 2.** Top 10 candidates generated and predicted with high ZT values based on the BERT based model, and Top 10 candidates generated and predicted with low ZT values based on the BERT based model.

| Prompt | Predicted ZT |
|---|---|
| Composition: {'Co': 1, 'Sb': 3, 'Ce': 0.078125, 'In': 0.03125, 'Ba': 0.03125}, Temperature: 682K | 1.7018 |
| Composition: {'Co': 1, 'Sb': 3, 'Ce': 0.171875, 'In': 0.03125}, Temperature: 534K | 1.6762 |
| Composition: {'Co': 1, 'Sb': 3, 'La': 0.140625, 'In': 0.015625, 'Ce': 0.046875}, Temperature: 558K | 1.5988 |
| Composition: {'Co': 1, 'Sb': 3, 'Ba': 0.0625, 'Ce': 0.046875, 'Y': 0.109375}, Temperature: 374K | 1.3705 |
| Composition: {'Co': 1, 'Sb': 3, 'La': 0.078125, 'Ce': 0.03125}, Temperature: 469K | 1.3429 |
| Composition: {'Co': 1, 'Sb': 3, 'In': 0.078125, 'Cu': 0.078125, 'Zr': 0.046875}, Temperature: 791K | 1.3407 |
| Composition: {'Co': 1, 'Sb': 3, 'Ba': 0.078125, 'W': 0.09375, 'La': 0.078125}, Temperature: 681K | 1.3268 |
| Composition: {'Co': 1, 'Sb': 3, 'Mn': 0.125, 'Yb': 0.078125, 'Cu': 0.03125}, Temperature: 386K | 1.3079 |
| Composition: {'Co': 1, 'Sb': 3, 'Ce': 0.03125, 'Li': 0.1875, 'Mg': 0.03125}, Temperature: 438K | 1.2964 |
| Composition: {'Co': 1, 'Sb': 3, 'Pb': 0.078125, 'Yb': 0.140625}, Temperature: 418K | 1.1031 |
| Composition: {'Co': 1, 'Sb': 3, 'Ag': 0.15625}, Temperature: 529K | 0.0717 |
| Composition: {'Co': 1, 'Sb': 3, 'Hf': 0.21875}, Temperature: 511K | 0.0862 |

| Composition: {'Co': 1, 'Sb': 3, 'Nd': 0.21875}, Temperature: 341K | 0.0863 |
| --- | --- |
| Composition: {'Co': 1, 'Sb': 3, 'Ag': 0.046875}, Temperature: 421K | 0.0989 |
| Composition: {'Co': 1, 'Sb': 3, 'Zn': 0.140625}, Temperature: 319K | 0.1042 |
| Composition: {'Co': 1, 'Sb': 3, 'Pd': 0.15625}, Temperature: 323K | 0.1102 |
| Composition: {'Co': 1, 'Sb': 3, 'Bi': 0.234375}, Temperature: 533K | 0.1121 |
| Composition: {'Co': 1, 'Sb': 3, 'Y': 0.03125}, Temperature: 539K | 0.1160 |
| Composition: {'Co': 1, 'Sb': 3, 'K': 0.125}, Temperature: 334K | 0.1160 |
| Composition: {'Co': 1, 'Sb': 3, 'C': 0.125}, Temperature: 550K | 0.1203 |

To evaluate the predictive reliability of the BERT-based model across the full range of ZT values, we generate 100 random candidate compositions with predicted ZT values greater than 1. The top 10 candidates with highest predicted ZT values are listed in Table 2. In parallel, we identify the 10 low-performing candidates with predicted ZT values less than 0.5, providing a spectrum-wide evaluation, also shown in Table 2. This dual-end analysis enables assessment of the model's validity not only in identifying promising thermoelectric materials but also in accurately flagging less favorable candidates. This contrastive evaluation further demonstrates the robustness and generalizability of the model's predictions.

## 3.3 Electrical and Thermal Transport

To validate our predictions, we select two compounds from Table 2: $CoSb_3Ag_{0.03125}$ and $CoSb_3Ce_{0.078125}In_{0.03125}Ba_{0.03125}$ which have the minimum and maximum predicted ZT values, respectively.

The pristine $CoSb_3$ model consists of a 2 × 2 × 2 supercell containing 256 atoms, which provides a sufficiently large configuration space with multiple intrinsic void sites suitable for subsequent substitution. Based on this supercell, two additional compositions were constructed: $CoSb_3Ag_{0.03125}$ was obtained by introducing one Ag atom into a void site, yielding a 257-atom structure. $CoSb_3Ce_{0.078125}In_{0.03125}Ba_{0.03125}$ was generated by introducing five Ce atoms, two In atoms, and two Ba atoms into the available void positions, resulting in a total of 265 atoms. Each structure was then fully relaxed using DFT calculations in VASP, as detailed in Section 2.3. The optimized geometries obtained from these calculations were obtained from CONTCAR files and used as the starting point for all subsequent electronic and thermal transport analyses. Images of these models are shown in Figure 4 (a)-(c).

The calculated electrical conductivity of pure $CoSb_3$ at 682 K, obtained using the Kubo–Greenwood formalism implemented in the *kg4vasp* package is $0.45 \times 10^5$ S/m, which is comparable to the experimental value of approximately $0.2 \times 10^5$ S/m reported in the literature [73]. The results from *kg4vasp* generally tend to overestimate the electrical conductivity or equivalently underestimate electrical resistivity, in a range of systems including Cu-Sb-S compounds and metallic aluminum suboxides. Our findings are consistent with this known trend[58,74]. Nevertheless, this level of agreement is sufficient for our purpose, which is to make a qualitative comparison of thermoelectric properties among the three systems and to estimate the ZT values qualitatively. For undoped $CoSb_3$, the electrical conductivity increases with temperature because of the thermal activation of charge carriers across the band gap, reflecting its semiconducting nature as reported in [75]. This behavior is also consistent with the electronic structure shown in Fig. 4(d), where the density of states exhibits a clear band gap.

**Table 3:** Electrical conductivity values for the three $CoSb_3$-based skutterudite systems.

| Thermo-electric material | Electrical conductivity ($10^5$ S/m) |
|---|---|
| $CoSb_3$ | 0.45 |
| $CoSb_3Ag_{0.03125}$ | 0.048 |
| $CoSb_3Ce_{0.078125}In_{0.03125}Ba_{0.03125}$ | 6.18 |

Table 3 summarizes the electrical conductivity values of the three $CoSb_3$-based skutterudite systems. The results clearly shows that Ce-In-Ba doped $CoSb_3$ has a significantly higher electrical conductivity than the undoped $CoSb_3$ and Ag-doped $CoSb_3$. Moreover, a qualitative comparison between electrical conductivity and ZT suggests a consistent trend: the LLM predicted the lowest ZT value for Ag-doped $CoSb_3$ and the highest ZT value for Ce-In-Ba doped $CoSb_3$.

To understand the enhanced or reduced electrical conductivity of the doped $CoSb_3$ systems relative to pure $CoSb_3$, Fig. 4 (d)-(f) shows the density of states (DOS) for the three systems. The DOS near the Fermi level is noticeably higher for $CoSb_3Ce_{0.078125}In_{0.03125}Ba_{0.03125}$ compared with pure $CoSb_3$ and Ag-doped $CoSb_3$. From the DOS profiles, the states in the vicinity of Fermi-level energy ($E_F$) appear to be on the order of $10^2$ states/eV/cell, whereas the corresponding values for the undoped and Ag-doped systems are substantially smaller. This higher DOS near the Fermi level is consistent with its significantly higher electrical conductivity.

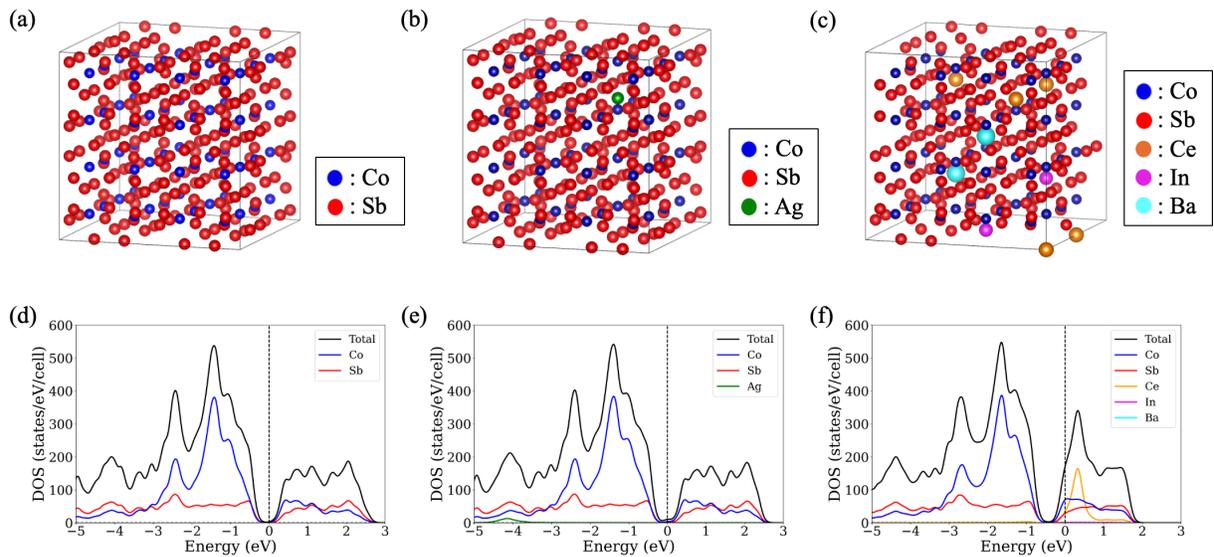

**Figure 4:** Atomic structures and electronic density of states (DOS) of $CoSb_3$-based skutterudites. (a–c) 2×2×2 supercell models of (a) pristine $CoSb_3$, (b) Ag-filled $CoSb_3$, and (c) the multi-filled $CoSb_3Ce_{0.078125}In_{0.03125}Ba_{0.03125}$ composition, where filler atoms occupy intrinsic void sites in the skutterudite framework. (d–f) The corresponding total and partial DOS plots show the evolution of electronic states near the Fermi level, indicated by the vertical dashed line.

To further support the predicted electrical conductivity values, we examine compounds similar to $CoSb_3Ce_{0.078125}In_{0.03125}Ba_{0.03125}$ that have been reported in the literature. For example, a Ba and In-based $CoSb_3$ skutterudite with the composition $Ba_{0.3}In_{0.2}Co_{3.95}Ni_{0.05}Sb_{12}$ exhibits an electrical conductivity value of $1.57 \times 10^5$ S m$^{-1}$, which is of the same order of magnitude that we calculated for the Ce-In-Ba doped system[76]. Another compound Ce-In-doped $CoSb_3$ skutterudite, $Ce_{0.2}In_{0.2}Co_4Sb_{12}$, has a high electrical conductivity of $1.4 \times 10^5$ S m$^{-1}$ and a ZT value of 1.4 [77]. These reported compositions support the plausibility of the high ZT value predicted by our model for the Ce-In-Ba doped skutterudite compound. Similarly, Ag-doped $CoSb_3$-based compounds reported in the literature [78–80] show relatively low electrical conductivity, around $0.5 \times 10^4$ S m$^{-1}$, consistent with our DFT results, as well as low ZT value of around 0.1[78], in agreement with the predictions of our LLM-based model. However, the lattice thermal conductivity of the three systems must also be examined, because thermal conductivity is another key factor contributing to the ZT value of a thermoelectric material.

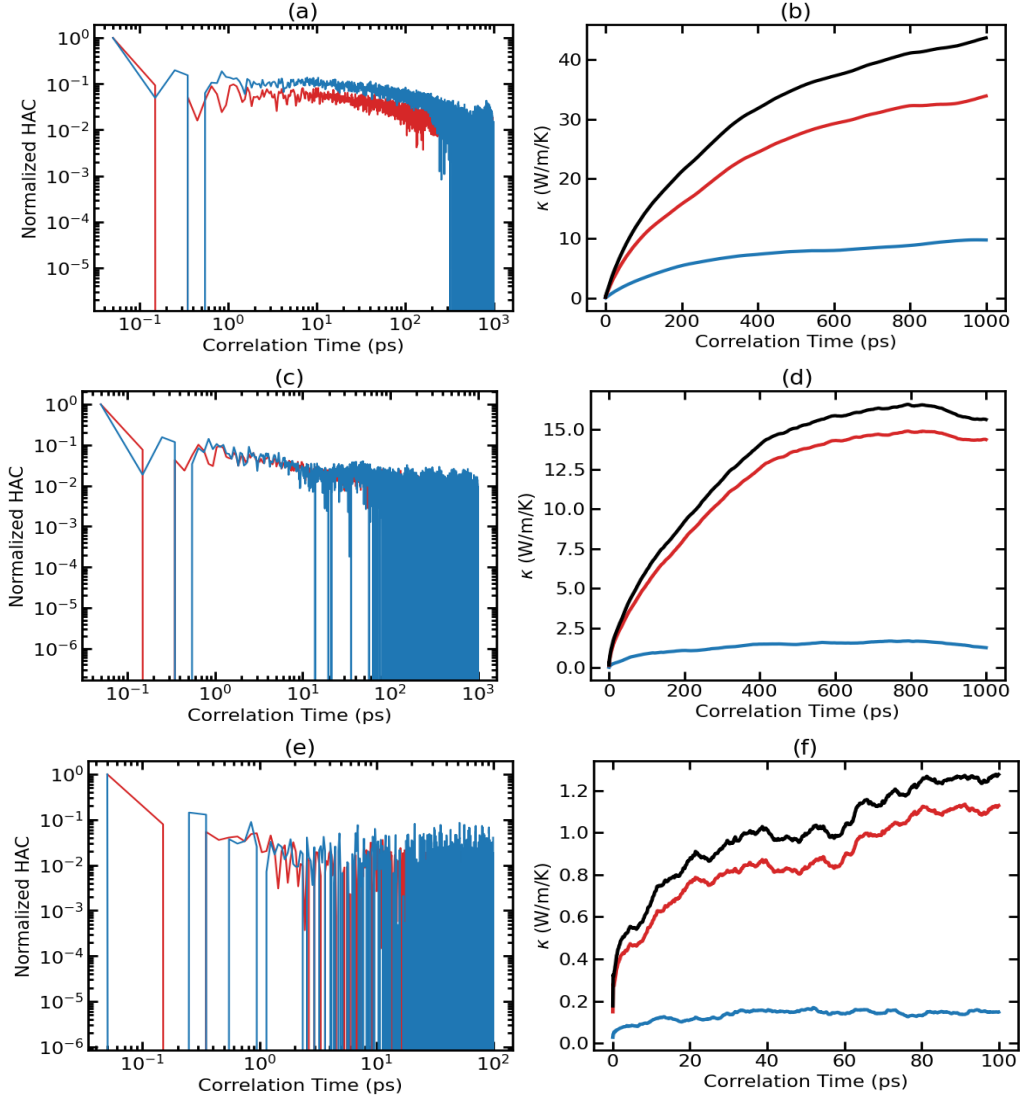

**Figure 5:** Normalized heat current autocorrelation (HAC) decay and the corresponding running lattice thermal conductivity obtained from equilibrium molecular dynamics simulations for $CoSb_3$-based skutterudites at 100 K. Panels (a)-(b) show the results for pristine $CoSb_3$, (c)-(d) for Ag-filled $CoSb_3$, and (e)-(f) for the Ce–In–Ba multi-filled $CoSb_3$ system. The left panels display the normalized HAC as a function of correlation time, illustrating the decay behavior of the heat current autocorrelation signal. The right panels show the running thermal conductivity extracted from the integrated HAC, where the blue curve represents the out-of-plane component ($\kappa^{out}$), the red curve represents the in-plane component ($\kappa^{in}$), and the black curve represents the total lattice thermal conductivity ($\kappa^{tot} = \kappa^{in} + \kappa^{out}$).

**Figures** 5 presents the normalized HAC as a function of correlation time, together with the in-plane, out-of-plane, and total lattice thermal conductivity extracted from the integrated HAC. The left panel plots Fig.5 (a), (c), and (e), shows how the HAC decays with increasing correlation time

on a logarithmic scale. The blue and the red curves represent the out-of-plane and in-plane contributions to the heat current autocorrelation function, respectively. A slower decay with correlation time indicates that the heat current remains correlated for longer, which corresponds to higher thermal conductivity. The right-panel plots, Fig. 5(b), (d), and (f), show the running thermal conductivity values of the three systems extracted from the integrated HAC. Based on these plots we conclude that the undoped $CoSb_3$ exhibits the slowest decay and therefore the highest thermal conductivity, whereas the Ce-In-Ba-doped $CoSb_3$ shows the fastest decay and the lowest thermal conductivity among the three systems. At temperatures near the predicted operating range of 600-700 K, the GPUMD simulations exhibit numerical instability, which is observed as non-physical values (NaNs) in the HAC function. This behavior is likely associated with limitations of the fitted FCP in accurately capturing high-temperature atomic vibrations and anharmonic effects. To ensure stable MD trajectories and well-converged HAC behavior, all simulations are therefore performed at 100 K. At this temperature, the HAC exhibited smooth decay, and the running thermal conductivity reached a stable plateau, enabling a consistent and reliable comparative analysis across the three systems. For undoped-$CoSb_3$, the calculated lattice thermal conductivity is approximately 40 W/(m·K) which is comparable to the experimental values reported by Guo et al. [81]. For the other two systems, the lattice thermal conductivity is about 15 W/(m·K) for Ag-doped $CoSb_3$ and about 1.2 W/(m·K) for $CoSb_3Ce_{0.078125}In_{0.03125}Ba_{0.03125}$. Qualitatively, the thermal conductivity of the $CoSb_3Ce_{0.078125}In_{0.03125}Ba_{0.03125}$ system is the lowest, which is consistent with its highest ZT value as predicted by our LLM model.

Similar to the discussion of electrical conductivity, we examine the thermal conductivity of compounds related to the Ce-In-Ba-doped system reported in the literature [76,77]. The compound $Ba_{0.3}In_{0.2}Co_{3.95}Ni_{0.05}Sb_{12}$ [76] exhibits a thermal conductivity value of 3.4 W/(m·K), while the compound $Ce_{0.2}In_{0.2}Co_4Sb_{12}$ [77] has thermal conductivity of 3 W/(m·K). These values are of the same order of magnitude as those obtained from our GPUMD simulation results for the Ce-In-Ba-doped $CoSb_3$ system. In contrast, the Ag-doped system, reported in the literature[78] has a thermal conductivity value of 5 W/(m·K), which is higher than that of the Ce-In and Ba-In-doped systems. Combined with the previously discussed electrical conductivity results, this trend suggests that multi-filled skutterudites such as Ce-In-Ba-doped systems, can achieve higher ZT values because of the favorable combination of high electrical conductivity and low thermal conductivity. In contrast, Ag-doped $CoSb_3$, characterized by lower electrical conductivity and higher thermal conductivity, exhibits a comparatively lower ZT value, consistent with our model predictions.

## 4. Conclusion

In this work, we demonstrate a workflow that couples large language model-based composition embeddings with classical neural networks, and first-principles-based transport calculations to accelerate the discovery and evaluation of $CoSb_3$-based skutterudites. By training a BERT-based regression model directly on experimental ZT value using only chemical composition and temperature data, we achieve substantially lower prediction errors compared to a conventional ANN model, enabling reliable screening of new filler combinations. The most promising candidates identified through this model are validated using DFT to estimate their electrical conductivity and through a Hiphive - GPUMD pipeline to evaluate their thermal conductivity, which together confirm the qualitative ZT trends predicted by the LLM approach. Our results show that the Ce–In–Ba filled composition exhibits enhanced electrical transport and significantly reduced lattice thermal conductivity relative to pristine $CoSb_3$ and Ag-filled $CoSb_3$, consistent with our LLM model's prediction of high thermoelectric performance. Overall, our work suggests that language-model embeddings can serve as practical, structure-free descriptors for quickly obtaining predictions of thermoelectric performance of different materials. In the future, this approach can be extended to material systems beyond skutterudites, and used for multi-property screening, for example, ZT values together with thermal and electrical conductivity, provided that suitable training datasets are available. This would help identify new thermoelectric materials with a wider range of potential applications.

**Acknowledgements**

This work was supported by the National Science Foundation (NSF) (Grant No. DMR- 2239216). This research used computational resources through the National Artificial Intelligence Research Resource (NAIRR) Pilot (NAIRR240231) and through the Sol cluster at Arizona State University (ASU). D. S. acknowledges The Fulton Undergraduate Research Initiative (FURI) at ASU.

All data are available on GitHub at: https://github.com/Yagnik12599/LLM_TE_MAT

**References:**

1. Brockway, P. E., Owen, A., Brand-Correa, L. I. & Hardt, L. Estimation of global final-stage energy-return-on-investment for fossil fuels with comparison to renewable energy sources. *Nat. Energy* **4**, 612–621 (2019).

2. Megia, P. J., Vizcaino, A. J., Calles, J. A. & Carrero, A. Hydrogen Production Technologies: From Fossil Fuels toward Renewable Sources. A Mini Review. *Energy and Fuels* **35**, 16403–16415 (2021).

3. Twaha, S., Zhu, J., Yan, Y. & Li, B. A comprehensive review of thermoelectric technology: Materials, applications, modelling and performance improvement. *Renewable and Sustainable Energy Reviews* **65**, 698–726 (2016).

4. Zhu, T. *et al.* High Efficiency Half-Heusler Thermoelectric Materials for Energy Harvesting. *Adv. Energy Mater.* **5**, 1500588 (2015).

5. Chen, G., Dresselhaus, M. S., Dresselhaus, G., Fleurial, J. P. & Caillat, T. Recent developments in thermoelectric materials. *International Materials Reviews* **48**, 45–66 (2003).

6. Snyder, G. J. & Toberer, E. S. Complex thermoelectric materials. *Nat. Mater.* **7**, 105–114 (2008).

7. Wood, C. Materials for thermoelectric energy conversion. *Reports on Progress in Physics* **51**, 459 (1988).

8. Zhang, X. & Zhao, L. D. Thermoelectric materials: Energy conversion between heat and electricity. *Journal of Materiomics* **1**, 92–105 (2015).

9. Itani, S., Zhang, Y. & Zang, J. Large Language Model-Driven Database for Thermoelectric Materials.

10. Takabatake, T., Suekuni, K., Nakayama, T. & Kaneshita, E. Phonon-glass electron-crystal thermoelectric clathrates: Experiments and theory. *Rev. Mod. Phys.* **86**, 669–716 (2014).

11. Nolas, G. S., Morelli, D. T. & Tritt, T. M. Skutterudites: a phonon-glass-electron crystal approach to advanced thermoelectric energy conversion applications. *Annual Review of Materials Science* **29**, 89–116 (1999).

12. Dolyniuk, J. A., Owens-Baird, B., Wang, J., Zaikina, J. V. & Kovnir, K. Clathrate thermoelectrics. *Materials Science and Engineering: R: Reports* **108**, 1–46 (2016).

13. Nolas, G. S. N. G. S. Clathrate Thermoelectrics. in *Chemistry, Physics, and Materials Science of Thermoelectric Materials: Beyond Bismuth Telluride* (eds. Kanatzidis, M. G., Mahanti, S. D. & Hogan, T. P.) vol. 69 107–120 (Springer US, 2003).

14. Zhou, A., Liu, L., Zhai, P., Zhao, W. & Zhang, Q. Electronic structure and transport properties of single and double filled CoSb3 with atoms Ba, Yb and in. *J. Appl. Phys.* **109**, (2011).

15. Hermann, R. P. *et al.* Einstein oscillators in thallium filled antimony skutterudites. *Phys. Rev. Lett.* **90**, 135505 (2003).

16. Itahara, H., Sugiyama, J. & Tani, T. Enhancement of electrical conductivity in thermoelectric [Ca 2CoO3]0.62[CoO2] ceramics by texture improvement. *Japanese Journal of Applied Physics, Part 1: Regular Papers and Short Notes and Review Papers* **43**, 5134–5139 (2004).

17. Li, L., Hu, B., Liu, Q., Shi, X. L. & Chen, Z. G. High-Performance AgSbTe2 Thermoelectrics: Advances, Challenges, and Perspectives. *Advanced Materials* **36**, 2409275 (2024).

18. Lamberton, G. A. *et al.* High figure of merit in Eu-filled CoSb3-based skutterudites. *Appl. Phys. Lett.* **80**, 598–600 (2002).

19. Harnwunggmoung, A. *et al.* Thermoelectric properties of Ga-added CoSb3 based skutterudites. *J. Appl. Phys.* **110**, (2011).

20. Popescu, B., Galatanu, M., Enculescu, M. & Galatanu, A. The inclusion of ceramic carbides dispersion in In and Yb filled CoSb3 and their effect on the thermoelectric performance. *J. Alloys Compd.* **893**, 162400 (2022).

21. Choi, S. *et al.* Enhanced thermoelectric properties of Ga and In Co-added CoSb3-based skutterudites with optimized chemical composition and microstructure. *AIP Adv.* **6**, 125015 (2016).

22. Ballikaya, S. & Uher, C. Enhanced thermoelectric performance of optimized Ba, Yb filled and Fe substituted skutterudite compounds. *J. Alloys Compd.* **585**, 168–172 (2014).

23. Yang, J. *et al.* Gadolinium filled CoSb3: High pressure synthesis and thermoelectric properties. *Mater. Lett.* **98**, 171–173 (2013).

24. Sales, B. C. Electron Crystals and Phonon Glasses: A New Path to Improved Thermoelectric Materials. *MRS Bull.* **23**, 15–21 (1998).

25. Sales, B., Mandrus, D., Chakoumakos, B., Keppens, V. & Thompson, J. Filled skutterudite antimonides: Electron crystals and phonon glasses. *Phys. Rev. B* **56**, 15081 (1997).

26. Hosseini Khorasani, S. A., Borhani, E., Yousefieh, M. & Janghorbani, A. Towards tailored thermoelectric materials: An artificial intelligence-powered approach to material design. *Physica B Condens. Matter* **685**, 415946 (2024).

27. Li, M., Dai, L. & Hu, Y. Machine Learning for Harnessing Thermal Energy: From Materials Discovery to System Optimization. *ACS Energy Lett.* **7**, 3204–3226 (2022).

28. Namsani, S., Gahtori, B., Auluck, S. & Singh, J. K. An interaction potential to study the thermal structure evolution of a thermoelectric material: β-Cu2Se. *J. Comput. Chem.* **38**, 2161–2170 (2017).

29. Jain, A., Shin, Y. & Persson, K. A. Computational predictions of energy materials using density functional theory. *Nature Reviews Materials 2016 1:1* **1**, 1–13 (2016).

30. Chelikowsky, J. R., Alemany, M. M. G., Chan, T. L. & Dalpian, G. M. Computational studies of doped nanostructures. *Reports on Progress in Physics* **74**, 046501 (2011).

31. Prashun, G., Vladan, S. & Eric, S. T. Computationally guided discovery of thermoelectric materials. *Nature Reviews Materials 2017 2:9* **2**, 1–16 (2017).

32. Sarikurt, S., Kocabaş, T. & Sevik, C. High-throughput computational screening of 2D materials for thermoelectrics. *J. Mater. Chem. A Mater.* **8**, 19674–19683 (2020).

33. Deng, T. *et al.* High-Throughput Strategies in the Discovery of Thermoelectric Materials. *Advanced Materials* **36**, 2311278 (2024).

34. Jia, X. *et al.* Unsupervised machine learning for discovery of promising half-Heusler thermoelectric materials. *NPJ Comput. Mater.* **8**, 1–9 (2022).

35. Na, G. S., Jang, S. & Chang, H. Predicting thermoelectric properties from chemical formula with explicitly identifying dopant effects. *NPJ Comput. Mater.* **7**, 1–11 (2021).

36. Kokyay, S. *et al.* A prediction model of artificial neural networks in development of thermoelectric materials with innovative approaches. *Engineering Science and Technology, an International Journal* **23**, 1476–1485 (2020).

37. Yang, K. T. Artificial Neural Networks (ANNs): A new paradigm for thermal science and engineering. *J. Heat Transfer* **130**, (2008).

38. Grossi, E. & Buscema, M. Introduction to artificial neural networks. *Eur. J. Gastroenterol. Hepatol.* **19**, 1046–1054 (2007).

39. Alrebdi, T. A. *et al.* Predicting the thermal conductivity of Bi2Te3-based thermoelectric energy materials: A machine learning approach. *International Journal of Thermal Sciences* **181**, 107784 (2022).

40. Naveed, H. *et al.* A Comprehensive Overview of Large Language Models. *ACM Trans. Intell. Syst. Technol.* **16**, (2025).

41. Openai, A. R., Openai, K. N., Openai, T. S. & Openai, I. S. Improving language understanding by generative pre-training. https://www.mikecaptain.com/resources/pdf/GPT-1.pdf (2018).

42. Vaswani, A. *et al.* Attention is all you need. *proceedings.neurips.cc* https://proceedings.neurips.cc/paper/2017/hash/3f5ee243547dee91fbd053c1c4a845aa-Abstract.html.

43. Chernyavskiy, A., Ilvovsky, D. & Nakov, P. Transformers: "The End of History" for Natural Language Processing? *Lecture Notes in Computer Science (including subseries Lecture Notes in Artificial Intelligence and Lecture Notes in Bioinformatics)* **12977 LNAI**, 677–693 (2021).

44. Choudhary, K. DiffractGPT: Atomic Structure Determination from X-ray Diffraction Patterns Using a Generative Pretrained Transformer. *Journal of Physical Chemistry Letters* **16**, 2110–2119 (2025).

45. Taori, R. *et al.* Stanford Alpaca: An Instruction-following LLaMA model. Preprint at https://github.com/tatsu-lab/stanford_alpaca (2023).

46. Wang, Y. *et al.* Self-Instruct: Aligning Language Models with Self-Generated Instructions. *Proceedings of the Annual Meeting of the Association for Computational Linguistics* **1**, 13484–13508 (2022).

47. Touvron, H. *et al.* LLaMA: Open and Efficient Foundation Language Models. https://arxiv.org/pdf/2302.13971 (2023).

48. Devlin, J., Chang, M. W., Lee, K. & Toutanova, K. BERT: Pre-training of Deep Bidirectional Transformers for Language Understanding. *NAACL HLT 2019 - 2019 Conference of the North American Chapter of the Association for Computational Linguistics: Human Language Technologies - Proceedings of the Conference* **1**, 4171–4186 (2018).

49. Chen, P. C. *et al.* A Simple and Effective Positional Encoding for Transformers. *EMNLP 2021 - 2021 Conference on Empirical Methods in Natural Language Processing, Proceedings* 2974–2988 (2021) doi:10.18653/V1/2021.EMNLP-MAIN.236.

50. Paszke, A. *et al.* PyTorch: An Imperative Style, High-Performance Deep Learning Library. https://doi.org/10.5555/3454287 (2019) doi:10.5555/3454287.

51. Wolf, T. *et al.* Transformers: State-of-the-Art Natural Language Processing. *EMNLP 2020 - Conference on Empirical Methods in Natural Language Processing, Proceedings of Systems Demonstrations* 38–45 (2020) doi:10.18653/V1/2020.EMNLP-DEMOS.6.

52. Calderín, L., Karasiev, V. V. & Trickey, S. B. Kubo–Greenwood electrical conductivity formulation and implementation for projector augmented wave datasets. *Comput. Phys. Commun.* **221**, 118–142 (2017).

53. Kubo, R. Statistical-Mechanical Theory of Irreversible Processes. I. General Theory and Simple Applications to Magnetic and Conduction Problems. *https://doi.org/10.1143/JPSJ.12.570* **12**, 570–586 (2013).

54. Migdal, K. P., Zhakhovsky, V. V., Yanilkin, A. V., Petrov, Y. V. & Inogamov, N. A. Transport properties of liquid metals and semiconductors from molecular dynamics simulation with the Kubo-Greenwood formula. *Appl. Surf. Sci.* **478**, 818–830 (2019).

55. Allen, P. B. & Feldman, J. L. Thermal conductivity of disordered harmonic solids. *Phys. Rev. B* **48**, 12581 (1993).

56. Raghuraman, V., Wang, Y. & Widom, M. An investigation of high entropy alloy conductivity using first-principles calculations. *Appl. Phys. Lett.* **119**, (2021).

57. Cohen, M. H. Review of the theory of amorphous semiconductors. *J. Non. Cryst. Solids* **4**, 391–409 (1970).

58. Di Paola, C., Macheda, F., Laricchia, S., Weber, C. & Bonini, N. First-principles study of electronic transport and structural properties of Cu12Sb4S13 in its high-temperature phase. *Phys. Rev. Res.* **2**, 033055 (2020).

59. Kresse, G. & Furthmüller, J. Efficient iterative schemes for *ab initio* total-energy calculations using a plane-wave basis set. *Phys. Rev. B* **54**, 11169 (1996).

60. Kresse, G. & Hafner, J. *Ab initio* molecular-dynamics simulation of the liquid-metal–amorphous-semiconductor transition in germanium. *Phys. Rev. B* **49**, 14251 (1994).

61. Kresse, G. & Furthmüller, J. Efficiency of ab-initio total energy calculations for metals and semiconductors using a plane-wave basis set. *Comput. Mater. Sci.* **6**, 15–50 (1996).

62. Blöchl, P. E. Projector augmented-wave method. *Phys. Rev. B* **50**, 17953 (1994).

63. Kresse, G. & Joubert, D. From ultrasoft pseudopotentials to the projector augmented-wave method. *Phys. Rev. B* **59**, 1758 (1999).

64. Kresse, G. & Hafner, J. Norm-conserving and ultrasoft pseudopotentials for first-row and transition elements. *Journal of Physics: Condensed Matter* **6**, 8245 (1994).

65. Perdew, J. P., Burke, K. & Ernzerhof, M. Generalized Gradient Approximation Made Simple. *Phys. Rev. Lett.* **77**, 3865 (1996).

66. Brorsson, J. *et al.* Efficient Calculation of the Lattice Thermal Conductivity by Atomistic Simulations with Ab Initio Accuracy. *Adv. Theory Simul.* **5**, 2100217 (2022).

67. Fan, Z. *et al.* GPUMD: A package for constructing accurate machine-learned potentials and performing highly efficient atomistic simulations. *Journal of Chemical Physics* **157**, (2022).

68. Eriksson, F., Fransson, E. & Erhart, P. The Hiphive Package for the Extraction of High-Order Force Constants by Machine Learning. *Adv. Theory Simul.* **2**, 1800184 (2019).

69. Fan, Z. *et al.* Thermal conductivity decomposition in two-dimensional materials: Application to graphene. *Phys. Rev. B* **95**, 144309 (2017).

70. Kubo, R. Statistical-Mechanical Theory of Irreversible Processes. I. General Theory and Simple Applications to Magnetic and Conduction Problems. *https://doi.org/10.1143/JPSJ.12.570* **12**, 570–586 (2013).

71. Green, M. S. Markoff Random Processes and the Statistical Mechanics of Time-Dependent Phenomena. II. Irreversible Processes in Fluids. *J. Chem. Phys.* **22**, 398–413 (1954).

72. Lepri, S., Livi, R. & Politi, A. Thermal conduction in classical low-dimensional lattices. *Phys. Rep.* **377**, 1–80 (2003).

73. Kawaharada, Y., Kurosaki, K., Uno, M. & Yamanaka, S. Thermoelectric properties of CoSb3. *J. Alloys Compd.* **315**, 193–197 (2001).

74. Huang, T. *et al.* Metallic Aluminum Suboxides with Ultrahigh Electrical Conductivity at High Pressure. *Research* **2022**, (2022).

75. Xu, B. *et al.* Thermoelectric properties of n-type CoSb3 fabricated with high pressure sintering. *J. Alloys Compd.* **503**, 490–493 (2010).

76. Wei, P. *et al.* Excellent performance stability of Ba and In double-filled skutterudite thermoelectric materials. *Acta Mater.* **59**, 3244–3254 (2011).

77. Li, H., Tang, X., Zhang, Q. & Uher, C. High performance InxCeyCo4Sb12 thermoelectric materials with in situ forming nanostructured InSb phase. *Appl. Phys. Lett.* **94**, (2009).

78. Nieroda, P., Kutorasinski, K., Tobola, J. & Wojciechowski, K. T. Search for Resonant-Like Impurity in Ag-Doped CoSb3 Skutterudite: Theoretical and Experimental Study. *Journal of Electronic Materials 2013 43:6* **43**, 1681–1688 (2013).

79. Saleemi, M. *et al.* Fabrication of nanostructured bulk Cobalt Antimonide (CoSb3) based skutterudites via bottom-up synthesis. *MRS Online Proceedings Library 2013 1490:1* **1490**, 42–47 (2013).

80. Wei, M. *et al.* Enhanced Thermoelectric Performance of CoSb3 Thin Films by Ag and Ti Co-Doping. *Materials* **16**, 1271 (2023).

81. Guo, R., Wang, X. & Huang, B. Thermal conductivity of skutterudite CoSb3 from first principles: Substitution and nanoengineering effects. *Scientific Reports 2015 5:1* **5**, (2015).